\documentclass[prl,twocolumn,showpacs,superscriptaddress]{revtex4}
\usepackage{amsmath,amsfonts,amssymb}
\usepackage{epsfig}
\usepackage{dcolumn}
\usepackage{amsmath}
\begin{document}

\title{On the validity and breakdown of the Onsager symmetry in mesoscopic
conductors interacting with environments}

\author{David S\'anchez}
\affiliation{Departament de F\'{\i}sica, Universitat de les Illes Balears,
E-07122  Palma de Mallorca, Spain}
\author{Kicheon Kang}
\affiliation{Department of Physics, Chonnam National University, 
 Gwangju 500-757, Korea}

\date{\today}

\begin{abstract}
We investigate magnetic-field asymmetries in the
linear transport of a mesoscopic conductor interacting with its
environment. Interestingly, we find that
the interaction between the two systems
causes an asymmetry only when the environment is out of equilibrium. 
We elucidate our general result
with the help of a quantum dot capacitively coupled to
a quantum Hall conductor and discuss the asymmetry dependence
on the environment bias and induced dephasing.
\end{abstract}

\maketitle

More than two decades ago
it became clear \cite{but86} that the Onsager-Casimir
symmetry relations \cite{ons,cas} are crucial to understand
the transport properties of mesoscopic conductors.
These symmetries are fundamentally
a consequence of microreversibility
of the scattering matrix that describes the conductance
of a phase-coherent conductor, dictating
that the two-terminal linear conductance $G$ must
be symmetric under reversal of the external magnetic field $B$.
An interesting consequence is that the phase of the
Aharonov-Bohm conductance oscillations of a solid-state
interferometer with a quantum dot embedded in one of
its arms can take the values 0 or $\pi$ only,
thus leading to phase rigidity~\cite{yac95,yey95}.

The scattering approach to mesoscopic transport
assumes that the mesoscopic conductor preserves
the electron quantum phase while inelastic processes
giving rise to irreversibility take place only in the reservoirs
that feed and draw the current. Therefore, close to equilibrium
$G$ is a function of the transmission $T$
evaluated at the Fermi energy $E_F$ common to all terminals and
the Onsager symmetry implies $T(B)=T(-B)$.
However, the system inevitably interacts with the
environment which may give rise to irreversible processes. 
Theoretically, the Onsager symmetry has been proved to be
valid for an isolated conductor. Therefore, it is an interesting question 
whether
a conductor interacting with its environment still fulfills the
symmetry.  In this Letter, we predict that this interaction 
leads, in fact, to magnetic-field {\em asymmetries}
(or, briefly, magneto-asymmetries)
when the environment is driven out of equilibrium.
In addition, our theory confirms why magnetic-field symmetries
are preserved in previous experiments on two-terminal 
conductors even if they cannot avoid interactions with its environment.

Recent works \cite{san04,spi04} have shown that
magneto-asymmetries
arise in {\em nonlinear} mesoscopic transport, a fact
which has been observed experimentally
\cite{rik05,wei06,mar06,let06,zum06,ang07}.
In contrast, here we address the magneto-asymmetry of the {\em linear}
mesoscopic conductance when the environment is out of equilibrium.

Consider the model system sketched in Fig. 1.
System $C$ is a mesoscopic conductor coupled to reservoirs $L$ and $R$.
The environment, denoted as system $D$, is modeled as a second conductor 
in close proximity with system $C$. There exists a Coulomb interaction
coupling conductor and environment electrons but no particle exchange
is permitted between the two subsystems. Experimentally,
the environment can be a quantum point contact (QPC), a quantum Hall bar
or any other system whose electron states depend on the electronic
trajectory across the conductor. The environment can be driven
out of equilibrium with applying a bias between reservoirs
$X$ and $Y$. As a consequence, we must
consider scattering of two particles described by the following
(asymptotic) states: $|1\rangle=|L\rangle_C \otimes |X\rangle_D$,
$|2\rangle=|L\rangle_C\otimes |Y\rangle_D$,
$|3\rangle=|R\rangle_C\otimes |X\rangle_D$ and
$|4\rangle=|R\rangle_C\otimes |Y\rangle_D$.
$|\cdots\rangle_C$ and $|\cdots\rangle_D$ represent the electron
states at system $C$ and $D$, respectively.
\begin{figure}[t]
\centerline{
\epsfig{file=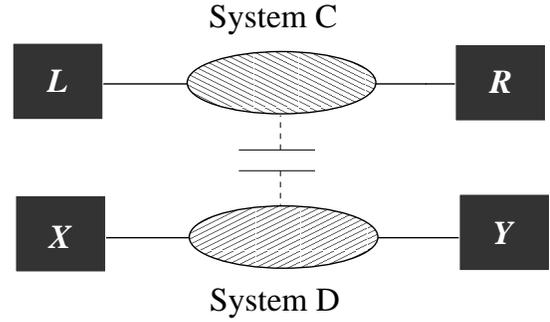,angle=0,width=0.4\textwidth,clip}
}
\caption{Schematic representation of the system under consideration.
System $C$ is a conductor capacitively coupled to an environment (system $D$).}
\label{fig1}
\end{figure}

Suppose that an electron from lead $L$ is injected into the conductor.
Before scattering, the initial density matrix $\rho_{\rm in}$ is given by
\begin{equation}
\rho_{\rm in}=(|L\rangle\langle L|)_C \otimes
(n_X|X\rangle\langle X| +n_Y |Y\rangle\langle Y|)_D\,,
\end{equation}
where $n_X$ and $n_Y$ are the transport electrons in system $D$
that originate from leads $X$ and $Y$, respectively.
Normalization condition, ${\rm Tr}\,\rho_{\rm in} = 1$, gives
the constraint $n_X+n_Y=1$.
At equilibrium, one has $n_X=n_Y=1/2$ whereas out of
equilibrium we rewrite $\rho_{\rm in}$ as
\begin{equation}\label{eq_rhoin}
\rho_{\rm in}=\frac{1}{2}(1+\delta\rho) |1\rangle\langle 1|
+ \frac{1}{2} (1-\delta\rho) |2\rangle\langle 2|
\,,
\end{equation}
where $\delta\rho=n_X-n_Y$ represents a nonequilibrium parameter.
At small bias, $\delta\rho$ is proportional to the bias voltage 
while very far from equilibrium
$\delta\rho=1$. The limit of $\delta\rho=1$ is discussed in 
Ref.~\onlinecite{khym06}. Upon scattering, the output density matrix
is \cite{taylor}
$\rho_{\rm out}=\hat{S}\rho_{\rm in}\hat{S}^\dagger$,
where $\hat{S}$ is the two-particle scattering matrix with
elements $S_{ij}=\langle i|\hat{S}|j\rangle$ ($i,j=1,\ldots,4$). 
Note that $\hat{S}$ describes scattering between two electrons in
different conductors~\cite{sun03}. 
We emphasize that $\hat{S}$ cannot, in general, be
written as the product of two single-particle scattering
matrices and that this nonseparability is precisely
due to the interaction between the two systems~\cite{note1}.

The transition operator, $\hat{T}$, gives the transmission probability
for an electron in system $C$, $T={\rm Tr}\,(\rho_{\rm out}\hat{T})$.
It reads
$\hat{T}=(|R\rangle\langle R|)_C \otimes I_D=
|3\rangle\langle 3|+|4\rangle\langle 4|$ ,
where $I_D$ is the unit operator in system $D$.
Then
we find
\begin{equation}\label{eq_t}
T=1-n_X (|S_{11}|^2+|S_{21}|^2)-n_Y(|S_{12}|^2+|S_{22}|^2)
\,,
\end{equation}
where unitarity of $\hat{S}$ has been used.

We introduce the magneto-asymmetry factor $\alpha\equiv T(B)-T(-B)$.
Microreversibility implies that $S_{ij}(B)=S_{ji}(-B)$. Hence,
we infer from Eq.~(\ref{eq_t}) that
\begin{equation}\label{eq_alpha}
\alpha=\delta\rho (|S_{12}|^2-|S_{21}|^2)
\,.
\end{equation}
Quite generally, one has $|S_{12}|^2\neq|S_{21}|^2$ and then the
transmission through the conductor is clearly not symmetric
under reversal of the magnetic field. 
Notably, the asymmetry
is proportional to $\delta\rho$, thus Eq.~(\ref{eq_alpha})
predicting that at small bias the magneto-asymmetry in the conductor
grows linearly with the voltage applied in system $D$.
Below we perform numerical simulations in a realistic system
that confirms this prediction. 

Our argument can be easily extended to the case where
the nonequilibrium situation of system $D$ includes
more than two leads. We note that Eq.~(\ref{eq_alpha}) is valid
when electrons are interacting within each subsystem and also
when the electron at system $C$ interacts with more than
one electron at $D$ but the problem then becomes involved
because one should resort to a multiple-particle scattering matrix.
Interestingly, the magneto-asymmetry vanishes 
when the environment is in equilibrium ($\delta\rho=0$), 
and it does not depend on a specific model for the environment.
This explains why all linear-transport experiments
satisfy the Onsager symmetry even though
the conductors cannot be isolated from uncontrolled interactions
with their environments.

It is clear from Eq.~(\ref{eq_alpha})
that the magneto-asymmetry is nonzero only to the
extent that $\hat{S}$ is nonseparable, i.e.,
$|S_{12}|^2=|S_{21}|^2$ if
$\hat{S}$ is given by the product $\hat{S}=\hat{S_C}\otimes \hat{S_D}$
where $\hat{S}_C$ and $\hat{S}_D$ are the single-particle scattering
matrices of system $C$ and $D$, respectively. Unitarity of $\hat{S}_D$ 
is necessary in deriving this relation. 
In Ref.~\cite{goo07}, $\hat{S}$ is expressed 
in terms of the scattering matrix of the uncoupled
systems to leading order in the interaction coupling strength 
and the correction term of the transmission probability
of the first system is found to depend on the injectivity of lead $X$ of the
second system. But the injectivity alone is {\em not} invariant under
field reversal~\cite{san04}. Therefore, $T$ need not be an even function
of $B$ due to interaction between the two systems.

We focus on the zero temperature case for simplicity
and assume that electrochemical potentials of leads
$X$ and $Y$ are $E_F+eV_D/2$ and $E_F-eV_D/2$
with $V_D$ the bias of system $D$.
In a two-dimensional conductor, 
$n_X\propto E_F+eV_D/2$ and $n_Y\propto E_F-eV_D/2$.
Using the constraint $n_X+n_Y=1$, we obtain
the nonequilibrium parameter $\delta\rho=n_X-n_Y=eV_D/2E_F$.
\begin{figure}[t]
\centerline{
\epsfig{file=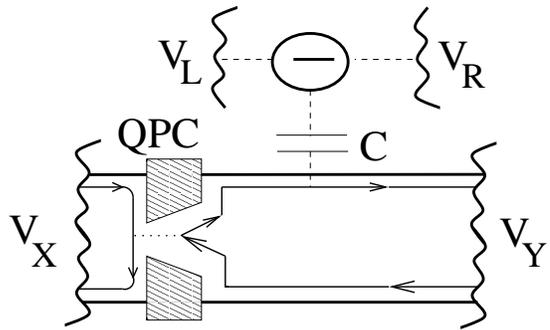,angle=0,width=0.4\textwidth,clip}
}
\caption{Sketch of a quantum dot capacitively coupled
with the top edge of a quantum Hall bar for $B>0$.
The edge state is transmitted at the QPC with
probability $T_d$. For $B<0$ the arrow directions
are reversed.}
\label{fig2}
\end{figure}

Let us now illustrate the general result given by Eq.~(\ref{eq_alpha})
with an instructive example as depicted in Fig. \ref{fig2}.
We consider resonant tunneling through a quantum dot
which is capacitively coupled to the top edge of
a quantum Hall conductor,
which works as the controllable environment.
We assume that the filling factor is $\nu=1$
and thus current is carried by two edge states
along opposite sides of the sample.

In addition, we consider a QPC constriction that partitions the
current injected from lead $X$ with probability $T_d$.
For the moment, consider the case $T_d=1$.
Qualitatively, it is clear that the current traversing the
dot, even close to equilibrium, is not an even function of $B$.
This follows from the fact that the potential of
the edge state $U_e$ is in equilibrium with the electrochemical potential
of the injecting lead. Then, $U_e=V_X$ for $B>0$
and $U_e=V_Y$ for $B<0$. Since the conductance through
the dot depends on the local potential at the dot $U_d$, and $U_d$,
in turn, depends on $U_e$ via the capacitive coupling, we must have,
quite generally, $G(B)\neq G(-B)$ \cite{note2}.

We investigate the case in which $U_d$ reacts continuously
to a change in $U_e$ and treat interactions in a 
mean-field way. We describe the scattering through the dot
with a Breit-Wigner resonance with level position $\varepsilon_0$ and
broadening $\Gamma=\Gamma_L+\Gamma_R$. 
$\Gamma_L$ ($\Gamma_R$) denotes
the resonance broadening contribution from the left (right) lead. 
To be definite, we take $\varepsilon_0$ and $\Gamma$ 
invariant under $B$ reversal, which is true at equilibrium.
To separate equilibrium and nonequilibrium contributions,
we consider the potential $U_d=U_{\rm eq}+\Delta U$.
Then, away from equilibrium the screening potential
$\Delta U$ follows from the charge neutrality condition which 
establishes that the net charge $\delta q=q_{\rm neq}-q_{\rm eq}$
must be equal to the polarization charge permitted by electrostatics:
\begin{equation}
\delta q(V_L,V_R,U_d)=
C (U_{\rm eq}+\Delta U-U_e) \,.\label{eq_c}
\end{equation}
The charge injected from the left and right electrodes,
$q_{\rm neq}=\int_{-\infty}^{E_F+eV_L} D_L(E-eU_{\rm eq}-e\Delta U)dE+ 
\int_{-\infty}^{E_F+eV_R} D_R(E-eU_{\rm eq}-e\Delta U)dE$,
is given by the injectivities $D_\alpha (E)=(e^2/2\pi) \Gamma_\alpha/|\Delta(E)|^2$ \cite{chr96},
where $\alpha=(L,R)$ and $\Delta(E)=E-\varepsilon_0+i\Gamma/2$.
The equilibrium charge, $q_{\rm eq}=\int_{-\infty}^{E_F} D_d(E-eU_{\rm eq})dE$,
depends on the total density of states, $D_d=D_L+D_R$.
In Eq.~(\ref{eq_c}), $C$ is the geometrical capacitance
between the edge state and the dot. If the density
of states of the edge state $D_e$ is much larger than $C/e^2$,
we simply have $U_e=V_X$ for $B>0$. In the general case,
we find
\begin{equation}\label{eq_ue}
U_e=\frac{C U_d+e^2D_e V_X}{C+e^2D_e} \,.
\end{equation}
for $B>0$. In the last equation, $V_X$ should be replaced with $V_Y$
for $B<0$. Upon inserting Eq.~(\ref{eq_ue}) in Eq.~(\ref{eq_c}),
we obtain $\delta q = C_\mu (U_\mathrm{eq} + \Delta U - V_X)$
where $C_\mu^{-1}=C^{-1}+(e^2D_e)^{-1}$ is
the {\em electrochemical} capacitance.
\begin{figure}[t]
\centerline{
\epsfig{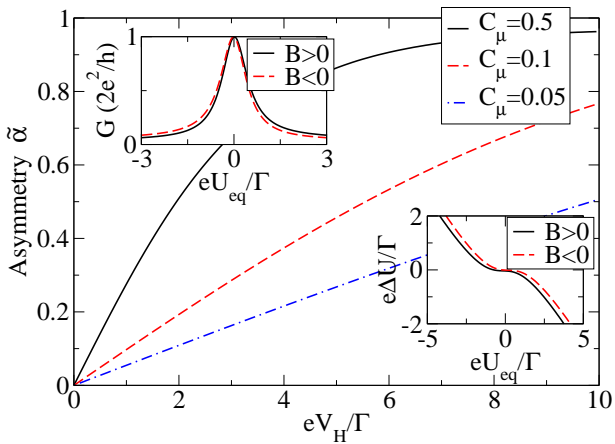}
}
\caption{(Color online) Magnetic-field asymmetry of the linear conductance
through the dot shown in Fig. \ref{fig2} for
$E_F=\varepsilon_0=0$, $\Gamma_L=\Gamma_R=\Gamma/2=0.1$ and $eU_{\rm eq}=\Gamma$
as a function of the voltage bias in the Hall bar $V_H$.
Upper inset: Linear conductance at different
polarizations of the magnetic field. Lower inset: Self-consistently
calculated screening potential. In the insets we take $C_\mu=0.1$
and $eV_H=\Gamma$.
}
\label{fig3}
\end{figure}

Equation~(\ref{eq_c}) is to be solved self-consistently.
Once we numerically find $\Delta U$, we can assess the current,
\begin{equation}\label{eq_cur}
I=\frac{2e}{h} \int_{E_F+eV_{R}}^{E_F+eV_L} \frac{\Gamma_L \Gamma_R}
{|\Delta(E-eU_{\rm eq}-e\Delta U)|^2} dE \,,
\end{equation}
and the linear conductance $G=dI/dV|_{V=0}$ with $V=V_L-V_R$.
Upper inset of Fig.~\ref{fig3} shows results for $G$
as a function of the equilibrium
level position $eU_{\rm eq}$ for a voltage bias
$-V_X=V_Y=V_H/2=\Gamma/2e$ applied in the Hall bar.
We observe that $G$ differs for opposite
$B$ orientations. The reason for the asymmetry
is uniquely due to the asymmetry
of the potential $\Delta U$ (see lower inset of  Fig.~\ref{fig3})
arising from the asymmetry of the Hall bar injectivity.
Thus, we present in Fig.~\ref{fig3} calculations of the
dimensionless magneto-asymmetry factor
$\tilde\alpha=[G(B)-G(-B)]/[G(B)+G(-B)]$
as a function of $V_H$ for
$eU_{\rm eq}=\Gamma$ and various $C_\mu$. 
These results are central to our discussion.
The magneto-asymmetry is larger for larger $C_\mu$
since the interaction coupling to the edge state is stronger.
Of course, in the limit $C_\mu\to 0$ (which amounts
to $C\to 0$), the magneto-asymmetry vanishes,
fulfilling the Onsager symmetry relation.
It also vanishes in the limit of an equilibrium environment
($V_H\to 0$). Moreover, $\tilde\alpha$ is linear with $V_H$
for small $V_H$, in excellent agreement with
Eq.~(\ref{eq_alpha}) \cite{note3}.
Saturation in $\tilde\alpha$ takes place for large $V_H$
and $C_\mu$, for which the precise
form of the local density of states
starts to play a role.

Let us now consider the case where the dot is coupled to
a partitioned edge state. There is a probability $T_d$ ($R_d=1-T_d$)
that the edge state is transmitted (reflected)
from lead $X$ to $Y$ ($X$) through the QPC.
For nonzero $C$, we find
\begin{equation}
U_e(B>0) = \frac{e^2 D_e (T_d V_X + R_d V_Y) + C U_d}{C + e^2 D_e}
\label{eq:Ue-partitioned}
\end{equation}
For $B<0$ one replaces $V_X$ with $V_Y$
in Eq.~(\ref{eq_ue}). As a result, $U_e$ is $B$-asymmetric
for nonzero $T_d$ (for $T_d=0$ the symmetry is restored) and depends
in a generic way on $T_d$ only for $B>0$.
%

More interestingly, this geometry gives rise to {\em magneto-asymmetry of
dephasing}. Dephasing in the dot is caused by current
partition in the Hall bar
\cite{spr00,sto99,lev00,but00,hac01,ave05,kang05}.
It is intimately related to the {\em possibility} of extracting
charge state information of the dot from the relative phase shift
between the transmitted and the reflected beam at the QPC. For $B>0$,  
Coulomb interaction induces dephasing because only the transmitted electron
undergoes a phase shift. For $B<0$, however, dephasing is not induced because
the dot interacts with the Hall bar before the electrons 
arrive at the QPC.
Therefore, it is clear that the magneto-asymmetry
of linear transport is caused not only by the asymmetry of the local
potential of Eq.~(\ref{eq:Ue-partitioned}) but also by the asymmetry of
dephasing.
\begin{figure}[t]
\centerline{
\epsfig{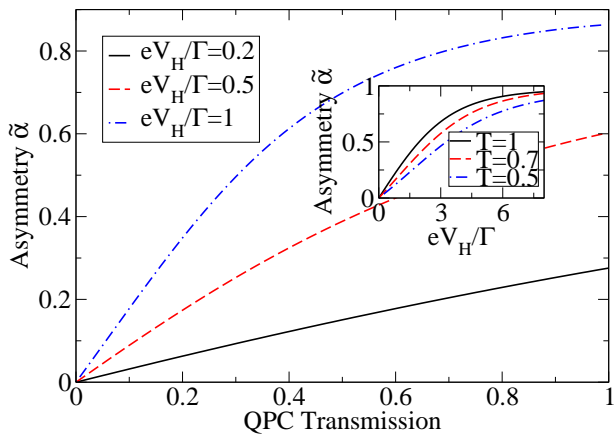}
}
\caption{(Color online) Magnetic field asymmetry as a function of the
QPC transmission for various biases applied
to the Hall bar $V_H$. Parameters are
$E_F=\varepsilon_0=0$, $\Gamma_L=\Gamma_R=\Gamma/2=0.1$,
$C_\mu=0.5$, and $eU_{\rm eq}=\Gamma$.
Inset: Asymmetry versus $V_H$ for different
QPC transmissions.}
\label{fig4}
\end{figure}
 
Dephasing induces an additional broadening
for $B>0$: 
$\Gamma\to\Gamma+\Gamma_\phi$ with $\Gamma_\phi=\eta T_d R_d V_H$,
where $\eta$ is a constant which depends on the details
of the interaction between the edge state and the
dot~\cite{lev00,but00,hac01,ave05,kang05}
($\eta=0$ for $B<0$).
Since our goal is to offer a simple picture
of the effect, we adopt
the phenomenological voltage probe model \cite{but86b}.
It assumes a fictitious
voltage probe attached to the dot with coupling $\Gamma_\phi$.
The condition $I_\phi=0$ determines
the potential at the probe, $V_\phi$. Every carrier
which enters the probe, is reemitted into the dot
with a completely unrelated phase,
thereby giving rise to dephasing. Thus, we add a term
$\int_{-\infty}^{E_F+eV_\phi} D_\phi(E-eU_{\rm eq}-e\Delta U)dE$
in the left-hand side of Eq.~(\ref{eq_c}) and a current contribution
$(2e/h) \int_{E_F+eV_{\phi}}^{E_F+eV_L} \Gamma_L \Gamma_\phi/
|\Delta(E-eU_{\rm eq}-e\Delta U)|^2 dE$ to Eq.~(\ref{eq_cur}).
Figure~\ref{fig4} shows the effect of current partitioning
in $\tilde\alpha$. The amount of dephasing is tuned
with $T_d$.
(To emphasize the role of $T_d$ in the asymmetry we use
a small $\eta=10^{-3}$).
When $T_d=0$, $U_e=V_Y$
independently
of the $B$ direction. As a result, $\tilde\alpha=0$. When $T_d$ increases,
$\tilde\alpha$ enhances monotonically. For higher $V_H$,
$\tilde\alpha$ becomes larger in agreement with Fig. \ref{fig3}.
Finally, in the inset we plot $\tilde\alpha$ as a function of $V_H$ for
decreasing values of $T_d$, which also demonstrates that $\tilde\alpha$
vanishes for $V_H\rightarrow 0$.

In conclusion, the statement that the
two-terminal linear conductance must be symmetric under
reversal of the magnetic field is widely accepted
and has been exhaustively confirmed.
However, conductors inevitably interact with the external environment.
We have shown that a magnetic-field asymmetry
appears even in the linear response 
when the environment is out of equilibrium.
This situation can be realized with another
conductor in close proximity applying an electric bias
across it. Importantly, we predict that the asymmetry
depends on the two-particle scattering matrix.
We have examined a quantum dot coupled to a quantum Hall
bar and found that the asymmetry grows with the 
Hall bar bias and the interaction coupling strength.
It also leads to an asymmetry of dephasing when the
Hall current is partitioned.

We thank M. B\"uttiker for useful suggestions.
This work was supported by the Spanish MEC Grant No.\ FIS2005-02796,
the ``Ram\'on y Cajal'' program, Korea Research Foundation 
(KRF-2005-070-C00055, KRF-2006-331-C00116), and APCTP focus program.


\begin{thebibliography}{90}
\bibitem{note1}
If one considers that systems $C$ and $D$ form a single, larger system
attached to four leads, the multiterminal single-particle
scattering approach applied to nonlinear transport
leads to a magneto-asymmetric current term
proportional to the biases applied to $C$ and $D$.
However, quantum correlation
effects between the two subsystems are not included 
in this approach.
\bibitem{note3}
We plot $\tilde\alpha$ instead of $\alpha$ because $\alpha$ vanishes
at very large $V_H$ as the conductance itself vanishes
(the dot goes far away from resonance).
\end{thebibliography}
\end{document}